\documentclass[pdflatex, iicol, sn-aps]{sn-jnl}

\usepackage{graphicx}%
\usepackage{multirow}%
\usepackage{amsmath,amssymb,amsfonts}%
\usepackage{amsthm}%
\usepackage{mathrsfs}%
\usepackage[title]{appendix}%
\usepackage{xcolor}%
\usepackage{textcomp}%
\usepackage{manyfoot}%
\usepackage{booktabs}%
\usepackage{algorithm}%
\usepackage{algorithmicx}%
\usepackage{algpseudocode}%
\usepackage{listings}%

\theoremstyle{thmstyleone}%
\theoremstyle{thmstyletwo}%

\theoremstyle{thmstylethree}%

\begin{document}

\title[Article Title]{Cold source of atomic hydrogen for loading large magnetic traps}


\author*[1]{\fnm{Aleksei} \sur{Semakin}}\email{assema@utu.fi}

\author[1]{\fnm{Janne} \sur{Ahokas}}\email{jmiaho@utu.fi}

\author[1]{\fnm{Otto} \sur{Hanski}}\email{otolha@utu.fi}

\author[1]{\fnm{Slava} \sur{Dvornichenko}}\email{viacheslav.dvornichenko@utu.fi}

\author[1]{\fnm{Tom} \sur{Kiilerich}}\email{tom.c.kiilerich@utu.fi}

\author[2]{\fnm{François} \sur{Nez}}\email{francois.nez@lkb.upmc.fr}

\author[2]{\fnm{Pauline} \sur{Yzombard}}\email{pauline.yzombard@lkb.upmc.fr}

\author[3]{\fnm{Valery} \sur{Nesvizhevsky}}\email{nesvizhevsky@ill.eu}

\author[4]{\fnm{Eberhard} \sur{Widmann}}\email{Eberhard.Widmann@oeaw.ac.at}

\author[5]{\fnm{Paolo} \sur{Crivelli}}\email{Paolo.Crivelli@cern.ch}

\author[1]{\fnm{Sergey} \sur{Vasiliev}}\email{servas@utu.fi}

\affil[1]{\orgdiv{Department of Physics and Astronomy}, \orgname{University of Turku}, \orgaddress{ \city{Turku}, \postcode{20014}, \country{Finland}}}

\affil[2]{\orgdiv{Laboratoire Kastler Brossel}, \orgname{Sorbonne Université, CNRS, ENS-PSL Université, Collège de France}, \orgaddress{\city{Paris}, \postcode{75252},  \country{France}}}

\affil[3]{\orgdiv{Institut Max von Laue - Paul Langevin}, \orgaddress{\street{71 avenue des Martyrs}, \city{Grenoble}, \postcode{38042}, \country{France}}}

\affil[4]{\orgdiv{Stefan Meyer Institute for Subatomic Physics}, \orgname{Austrian Academy of Sciences}, \orgaddress{\street{Dominikanerbastei 16}, \city{Vienna}, \postcode{1010}, \country{Austria}}}

\affil[5]{\orgdiv{Institute for Particle Physics and Astrophysics}, \orgname{\textsc{eth}  Zurich}, \orgaddress{\city{Zurich}, \postcode{8093}, \country{Switzerland}}}


\abstract{We present a design and performance tests of an intense source of cold hydrogen atoms for loading large magnetic traps. Our source is based on a cryogenic dissociator of molecular hydrogen at 0.6~K followed by a series of thermal accommodators at 0.5, 0.2 and 0.13~K with inner surfaces covered by a superfluid helium film. All components are thermally anchored to corresponding stages of a dilution refrigerator. The source provides a continuous flux of $7\times10^{13}$~H~atoms/s in a temperature range of 130-200~mK. We have successfully used the source for loading a large Ioffe-Pritchard magnetic trap recently built in our laboratory [Ahokas et.al., Rev. Sci. Instr. \textbf{93}(2), 023201 (2022)]. 
Calorimetric measurements of the atomic recombination heat allow reliable determination of the atomic flux and H gas density in the trap. We have tested the performance of the source and loading of H atoms into the trap at various configurations of the trapping field, reducing the magnetic barrier height to 75$\%$ and 50$\%$ of the nominal value of 0.8~T~(0.54~K) as well as at the open configuration of the trap at its lower end, when the atoms are in contact with the trapping cell walls covered by a superfluid helium film. In the latter case, raising the trapping cell temperature to 200-250~mK, the low-field seeking atoms at densities exceeding 10$^{11}$~cm$^{-3}$ can be stored for the time over 10$^3$~s, sufficiently long for experiments on precision spectroscopy of cold H gas.}





\maketitle
\clearpage
\section{Introduction}
\label{sec:intro}
Experiments with hydrogen atoms were at the core of many discoveries in quantum physics and precision spectroscopy in the history of physics. 
The dispersion of atomic velocities due to the thermal motion is always a limitation for reaching highest resolution and accuracy. Obtaining an intense beam of slow hydrogen atoms or cooling trapped hydrogen gas is important for further improvements in optical, hyperfine and gravitational spectroscopy~\cite{comparat2024}  and for potential discoveries of new phenomena. Laser cooling routinely used for many other atoms and molecules is very difficult for hydrogen due its low mass and VUV range of the required excitation from the ground electronic state. Magnetic trapping and evaporative cooling are the methods which were successfully used for reaching quantum degeneracy and BEC in H~gas in the end of the last century~\cite{Fried1998}. Revisiting this techniques, we have recently built a large magnetic trap which is planned to be used for a variety of experiments with ultra-cold hydrogen atoms by the GRASIAN collaboration~\cite{ IPT_RSI}. Loading such a trap requires an intense source of low-field seeking~(\textit{lfs}) atoms at sub-Kelvin temperatures. Here, we describe the design and operation of such a source and demonstrate its efficiency for a loading the magnetic trap. The source is based on a cryogenic dissociator operating at 0.65~K and a hydrogen transfer line covered by superfluid helium film and thermally anchored to different stages of a dilution refrigerator. A gradient of magnetic field separates low-field and high-field seeking~(\textit{hfs}) atoms. The source can be used as a stand-alone beam of slow hydrogen atoms. Elastic scattering from a superfluid helium covered surfaces can be utilized for further manipulation of the beam velocity and intensity.

In Section~2 of this paper, we present a brief review of the cold atomic hydrogen sources and techniques used for various experiments, in Section~3 we describe shortly our experimental setup with a magnetic trap and a dilution refrigerator, Section~4 contains a description of the low-temperature hydrogen dissociator and hydrogen transfer line with the thermal accommodators. We present results of the tests of the source operation and loading the magnetic trap as well as of decay of the trapped gas in the Section~5. Finally, we analyze further improvements of the source for getting colder atoms with higher flux and suggest some ideas on the manipulation of the cold H~beam using quantum reflection from a superfluid helium covered surface.

\section{Background}
\label{sec:Backgr}
In the pioneering work of I.~Silvera and J.~Walraven~\cite{Walraven1982} an intense cryogenic beam of H was obtained by dissociating H$_2$ at room temperature in the RF discharge and further thermal accommodation in a nozzle cooled by liquid helium. The atoms were cooled to $\approx 8$~K in collisions with the nozzle walls and a fairly large flux of $\phi_H~\simeq~2.4~\times 10^{16}$~atoms/s was demonstrated. Such a cold nozzle technique was then used for high precision optical spectroscopy with atomic hydrogen~\cite{Parthey2011, Matveev2013, Brandt2022}, microwave spectroscopy~\cite{Diermaier2017} and experiments on quantum reflection of slow hydrogen beam~\cite{Killian2023, Killian2024}. In these experiments the typical thermal distribution of atomic velocities after the nozzle corresponded to the temperature of $\sim 6$~K with a velocity peak at $\sim350$~m/s. A special time-of-flight technique allowed a selection of slowest atoms in this distribution down to 50-60~m/s~\cite{Cooper2020, Killian2024}. Further reduction was not possible due to the low intensity of the atomic flux and insufficient sensitivity of the detection system. 

The performance of the cold nozzle technique is limited by the recombination of atoms adsorbed on its wall which is strongly enhanced at low temperatures. It turned out that the only way to go further down in temperature is to use a superfluid helium film for the wall coverage which has the lowest adsorption energy of H~$\sim 1$~K~\cite{Safonov2001, cline1981, morrow1981, matthey1981}. This has been realized in experiments by I.~Silvera and J.~Walraven~\cite{Silvera1980} where the atoms were stabilized in a sample cell covered by superfluid helium and located in a strong magnetic field of 7~T. The atoms were accumulated in a high-filed seeking~(\textit{hfs}) spin state which strongly suppressed recombination. In the follow up experiments H densities above $10^{17}~$~cm$^{-3}$ were reached at temperatures of 0.2-0.5~K~\cite{Cline1980magnetic, Hess1984}. 

At this stage of experiments with spin-polarized hydrogen, dissociation of molecular H$_2$ was performed at room temperature after which the atoms were transported into the 4~K nozzle and then in the low temperature region covered by a superfluid helium film. The superfluid film flow towards warmer regions, evaporating at $\sim 1$~K and re-condensing back onto the cold surfaces, created severe heat load on the dilution refrigerator. Also, the part of the H transfer line between the 4~K nozzle and the 1~K region, not covered by a superfluid helium film, had large losses due to the recombination. A much simpler method for producing cold H was introduced by W.~Hardy at UBC~\cite{Jochemsen1982, Statt1985}. They built a cryogenic H$_2$ dissociator in which dissociation of H$_2$ was performed in a cryogenic discharge below 1~K. This technique was later used by nearly all groups working with spin-polarized H. A typical flux of \textit{hfs} H into the sample cell located in a strong magnetic field was 1-5~$\times 10^{13}$~atoms/s at temperatures down to $\sim100$~mK. Although superfluid helium film was used to cover the walls, further cooling of H~gas was not possible due to the adsorption of the atoms on its surface which finally became an insurmountable obstacle.

To overcome the surface recombination problems and reaching BEC, a magnetic trapping of \textit{lfs} H atoms with evaporative cooling was proposed by H.~Hess~\cite{Hess1986} and realized in experiments by the MIT~\cite{Hess1986maser} and Amsterdam~\cite{VanRoijen1987} groups. Transport of the gas into the magnetic trap partially occurred in the low field region where the recombination and relaxation processes are substantially faster. Atomic fluxes of \textit{lfs} were somewhat lower, approaching $\sim 5\cdot10^{12}$~atoms/s~\cite{Hess1986maser, VanRoijen1987}, which was however sufficient for loading small magnetic traps at MIT and Amsterdam.

The above mentioned methods of cold~H gas production based on cryogenic techniques, require sophisticated equipment based on dilution refrigerators. Other approaches were pursued to avoid these complications, e.g. using all-optical systems for decelerating H~beams or using Zeeman decelerators (see ref.~\cite{Jansen2020} for a review). Laser deceleration methods are well known and used for a large variety of atoms and molecules. For a hydrogen, straightforward implementation of this techniques requires excitation of 1S-2P transition at the wavelength of 121.6~nm the which so far remains a challenge. Several other methods were proposed, including a two-photon deceleration using 1S-2S transition at 243~nm~\cite{Cooper2018}, an optical deceleration of hydrogen beam, using a moving optical lattice~\cite{Cooper2023} and a magic wavelength trapping of H~\cite{Udem2024}.

\section{Experimental setup for magnetic trapping of H gas}
One of the goals of the GRASIAN collaboration~\cite{vasiliev2019, GRASIAN} 
is to build a large magnetic trap for atomic hydrogen, accumulate large quantities of hydrogen gas in the trap, and cool the gas to temperatures below 1~mK. The atoms can be used in various experiments or will be transferred into a second shallower trap for a further cooling into the $\mu$K range. A magnetic trap of the Ioffe-Pritchard type~(IPT) was built recently in Turku~\cite{IPT_RSI}. The trap consists of an octupole magnet~(OM) for radial confinement, two pinch coils for axial confinement~(UP and LP), and a 3~T magnet used for a cryogenic dissociator~(DS). The atoms are confined by the magnetic field barrier of $\approx0.54$~K. Hydrogen gas is isolated from the vacuum space of the dilution refrigerator~(DR) in a plastic sample cell~(SC) located in the bore of the IPT and covered inside by superfluid helium film. The IPT and SC geometry is optimized for a largest effective volume of the trapped gas. The SC has an inner diameter of 10.5~cm and a length of 35~cm. Its walls are thermally anchored to the mixing chamber of the DR and can be cooled down to 120~mK.

Aiming at accumulation and storage of the largest possible amounts of H~gas, one could also increase the gas density. However, the decay of the trapped gas is governed by the process of two-body dipolar relaxation, with the rate increasing as a second power of density. At our conditions the rate constant $\approx 10^{-15}$~cm$^3$/s~\cite{Stoof1988} sets a limitation on the maximum density of the stored gas for a given lifetime, e.g. at the density of $10^{12}$~cm$^{-3}$ the half-decay time is $\sim1000$~s. Then, the total number of trapped atoms of $6\times 10^{14}$ is sufficiently large to start evaporative cooling and reach our goal: over~$10^{12}$ atoms below 1~mK. Realization of these plans requires a powerful enough source of atoms for filling the SC. For accumulating the above mentioned number of atoms within 10~minutes the incident flux should exceed $2\times10^{13}$~\textit{lfs}~atoms/s.

\begin{figure}
  \includegraphics[width=\columnwidth]{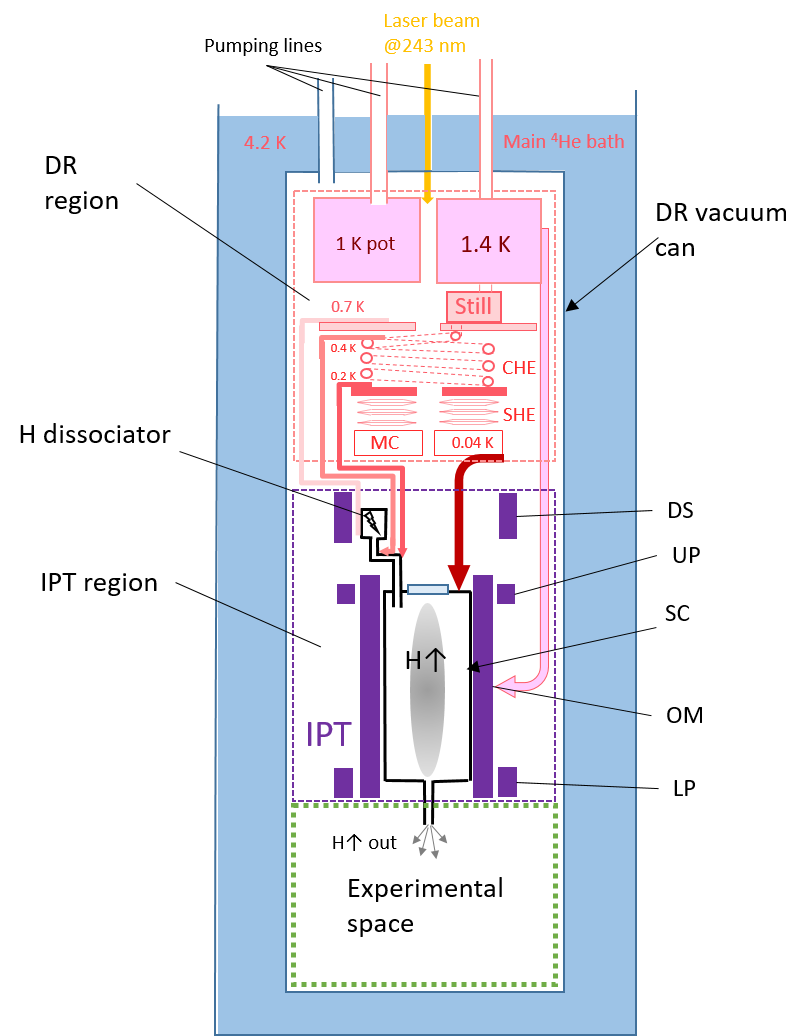}
  \caption{Schematic of the experimental setup based on a large Ioffe-Pritchard trap (IPT). DR  - dilution refrigerator, DS - dissociator solenoid, UP - upper pinch coil, SC - Sample cell, OM - octupole magnet, LP - lower pinch }
  \label{fig:ExpSetup}
\end{figure}

Our cryogenic setup (see Fig.~\ref{fig:ExpSetup}) is based on a wet-type dilution refrigerator Oxford~2000, with a cooling power of 300-500~$\mu$W at 100~mK. The IPT~\cite{IPT_RSI} is mounted inside the vacuum can of the refrigerator and thermally anchored to its 1~K~pot operating nominally at 1.4~K. An experimental space below the IPT is 40~cm high and 22~cm in diameter. We plan to arrange there a second shallower trap for further cooling the H gas into the sub-mK range. Experiments on quantum reflection of ultra-slow atoms from a liquid helium surface, observations of a Gravitational Quantum States above a liquid helium surface of finite thickness~\cite{Crépin_2017} and a superfluid He bulk~\cite{crepin2019}, and precision spectroscopy are on the list to be set up in this space.

\section{The H source}
\label{sec:source tech}
\subsection{General considerations}
Dissociation of molecular hydrogen is normally performed by electron impact in a plasma of DC or RF discharge in a gas. Maintaining a discharge continuously in a sufficiently large volume can not be done below 1~K. Therefore, the first cryogenic hydrogen dissociator constructed by W.~Hardy and coworkers~\cite{Jochemsen1982, Statt1985} operated in a pulsed mode with the duty cycle $\sim 10^{-3}$, typical energy per pulse $\sim10~\mu$J and average power of $\sim 1$~mW. This technique has been utilized by several other groups, with a most detailed study performed by Helffrich~\cite{Helffrich1987}. The inner surfaces of the dissociator chamber of several cm$^3$ volume were plated by a thin layer of solid H$_2$, on top of which a superfluid helium film was condensed. During the RF pulse, the helium film is partially evaporated and a discharge starts in the helium vapor. Electrons of the discharge bombard the solid H$_2$ layer and dissociate molecules. Hydrogen atoms partially remain trapped inside the solid layer, partially escape into the bulk of the chamber. After the RF pulse, helium condenses rapidly back onto the wall, and the remaining hydrogen gas flows from the dissociator into the experimental chamber. Applying a magnetic field in the region of the dissociator or in the experimental chamber below, provides separation of  \textit{hfs} from \textit{lfs} atoms and reduces recombination.  

One of the main advantages of the cryogenic dissociator operating below 1~K is that it provides a simpler and more reliable solution of the problem related with the re-condensing of the superfluid helium film. Compared with the cold nozzle followed by a liquid helium covered part of the transfer tube, the heat load on the refrigerator and the atomic recombination are reduced by orders of magnitude. The suppression of the recombination is especially effective for experiments with \textit{hfs} atoms. The dissociator is located at the fringe of the superconductive magnet and transport of the atoms into the sample cell is assisted with the magnetic field gradient. The sample cell may be cooled to temperatures approaching 150-200~mK without substantial atomic loss at the surface because of the high degree of polarization of the electron and the nuclear spin in the strong magnetic field~\cite{BlueBible}. 

The situation is different for loading magnetic traps with \textit{lfs} atoms. In this case, the magnetic field gradient is opposite: the dissociator is located in the strong field of 3-4~T and the \textit{lfs} atoms need to travel into the low field region of the magnetic trap, being cooled to the trappable temperature (typically $\sim$100~mK) by thermalization with the walls. Recombination on the surface at the final stage of the transfer line is much faster and the line design requires special care. In experiments in Amsterdam and MIT this was resolved by shortening the part between the dissociator and the trapping cell as much as possible which still allowed to cool the trapping cell to below 100~mK without discharge. During the discharge operation, the trapping cell was overheated to 200-300~mK. Continuous fluxes of \textit{lfs}  $\approx 2\times 10^{12}$~atoms/s at 150~mK were reported in the trapping experiments at Amsterdam~\cite{VanRoijen1987}. The MIT group used a short term operation mode for loading their magnetic trap: a discharge was run for 30~s during which the trapping cell was heated up to 300~mK~\cite{Killian1999}. The estimated flux of \textit{lfs} was $\sim 10^{13}~$~atoms/s. After loading, the trapping cell required about 3~minutes to cool to 100~mK. During this time, the gas was not isolated from the walls leading to a large recombination loss. 

In the construction of our H source described below, we aimed at a continuous operation with the H~gas flux above $10^{13}$~atoms/s and entering the SC at a temperature $\sim100$~mK. This required three thermal accommodation stages between the cryogenic dissociator and the trapping cell.

\subsection{Cryogenic dissociator}
Our cryogenic dissociator has a similar construction to that of the trapping experiments in Amsterdam~\cite{VanRoijen1987}. We built it and have used in our lab for nearly 20 years for studies of \textit{hfs} hydrogen in high magnetic field. It is a cylindrical copper chamber with the inner diameter of 1.8~cm and the length of 3~cm. A helical resonator made of 1.5~mm thick copper wire is located inside the chamber and inductively coupled by one loop to a coaxial cable feed-through at the top of the chamber~(Fig.~\ref{fig:Hsource_setup}). The helical resonator has a resonance frequency of 330~MHz and a quality factor of $\sim 200$. The dissociator chamber is thermally anchored to the bottom of the Still of the dilution refrigerator having typically a temperature $\approx 0.6$~K. The thermal link is provided via an annealed high purity copper rod of 12~mm diameter and length of 40~cm. At the top of the dissociator chamber there is a thermally isolated feedthrough of a capillary for loading H$_2$. Typically $\sim 1$~mmole of H$_2$ was condensed onto the inner surfaces of the dissociator chamber at $\sim1$~K while heating the inlet capillary and the rest of the H$_2$ loading line above 15~K. Loading process lasts typically $\sim30$ min and is controlled by the frequency shift of the helical resonator. Typically, after condensing H$_2$ the resonator frequency decreased by $\approx0.3$~MHz. This implies that part of the H$_2$ is also condensed on the surface of the helix, but does not provide reliable evaluation of the solid hydrogen layer thickness. From our previous experience such amount is sufficient for several weeks of continuous source operation below 1~K.

Normally we operate our cryogenic dissociator with RF pulses of 0.5-1~ms length and 20-50~Hz repetition rate. RF power is provided by an oscillator operating in CW mode which is then modulated by a rectangular pulse and finally amplified by a 10~W amplifier. We estimate that about 5/1~mW of peak/average power is absorbed in the dissociator when it operated in the optimal conditions for getting largest flux of atoms. This leads to the overheating of the dissociator by $\approx 40$~mK above the value when the discharge is off. The power load to the Still of the dilution refrigerator caused by the RF discharge is substantially lower than the $\sim10$~mW heating required to provide nominal circulation rate of the refrigerator.

\subsection{Hydrogen transfer line}
\begin{figure}
  \includegraphics[width=\columnwidth]{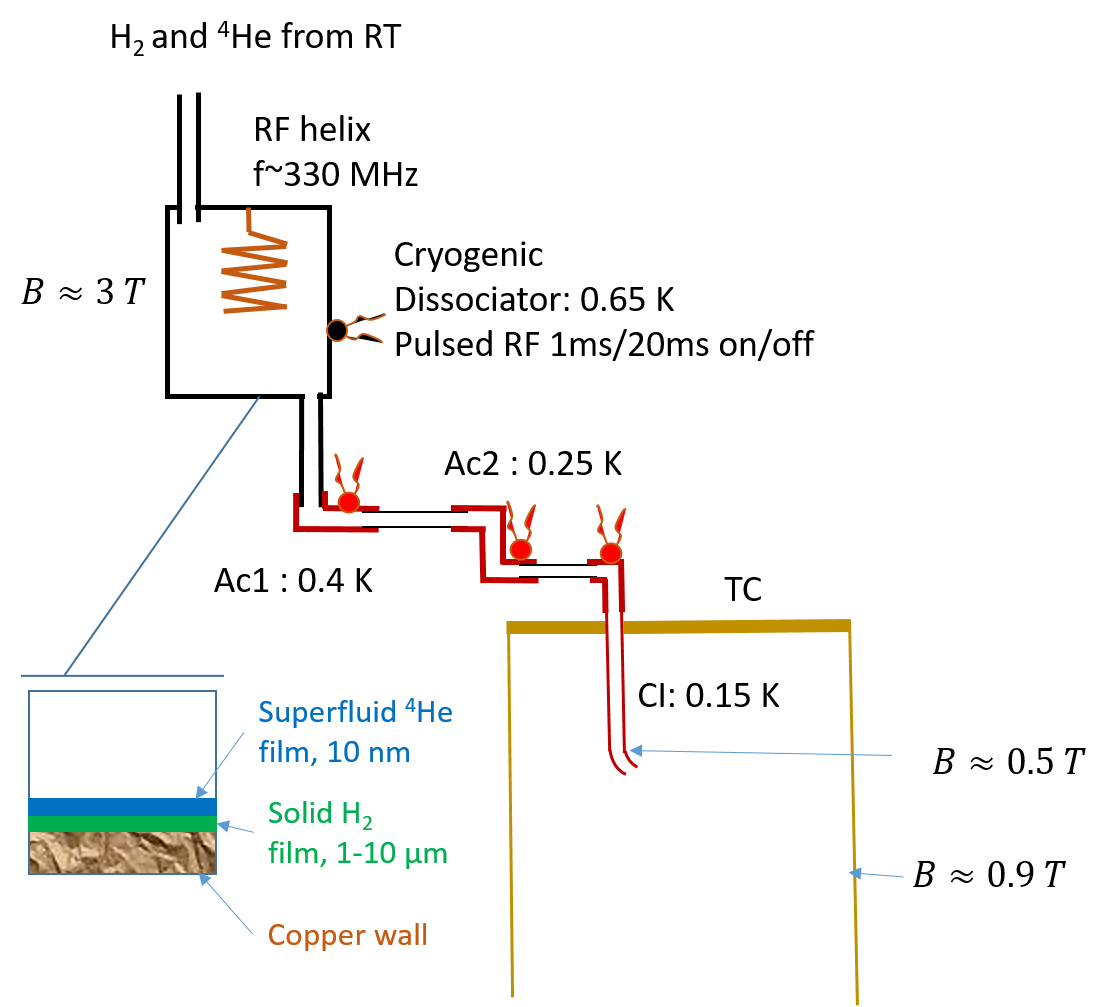}
  \caption{Schematic of the H source setup. Atomic hydrogen is generated in a cryogenic dissociator at $\approx$0.65~K, passes two thermal accommodation stages Ac1~(0.4~K) and Ac2~(0.25~K), and enters the SC via the cell inlet tube (CI) at 0.13~K. The lower end of the CI is extended approximately into the middle of the magnetic trapping well.}
  \label{fig:Hsource_setup}
\end{figure}

Hydrogen gas produced in the dissociator contains both electron spin components. For reducing recombination on the way to the trapping cell having temperature of $\sim100$~mK, these spin state should be separated as soon as possible after leaving the dissociator. Therefore, the transfer line is designed to provide a strong magnetic field gradient along its sections and an optimal temperature gradient for efficient thermal accommodation~(Fig.~\ref{fig:Hsource_setup}). After the dissociator, located in a 3~T magnetic field, the gas of \textit{lfs} is pulled into the first thermal accommodator~(Ac1) which is thermally anchored to an upper part of the continuous heat exchanger of the DR bringing the diluted stream of the mixture~($^3$He/$^4$He) into the Still. It has an unloaded temperature of $\approx 360$~mK. After a 90$^{\circ}$ turn the gas flows into the second accommodator~(Ac2) having unloaded an temperature of $\approx220$~mK. It has a thermal link to the 0.1~K plate of the DR which is cooled by the diluted stream leaving the upper step heat exchanger of the DR. Ac2 has two 90$^{\circ}$ turns. Both accommodators are machined out of a block of high purity annealed copper. The accommodators are connected to each other and to the dissociator by pieces of thin walled CuNi tubes of 4.5 mm i.d. and the length of $\approx 2.5$~cm. After the Ac2 the gas enters the last section called the cell inlet~(CI) via an In sealed flange and edge welded bellows. CI is thermally anchored to the mixing chamber~(MC) of the DR and can be cooled down to the minimum temperature of $\approx 130$~mK. 

Adding $^4$He to the H loading system creates a superfluid film on its walls. The film is pulled into the regions of higher temperature by the fountain pressure. Inside the dissociator the saturated vapor pressure of $^4$He is sufficiently high to create a stream of the vapor moving down and condensing back in the Ac1 and Ac2. This leads to an extra thermal link between the transfer lines sections and may change its temperatures significantly. To minimize this effect, strong thermal links to the corresponding parts of the DR were made using 12 mm diameter high purity copper rods. To verify the overheating effects due to the presence of a superfluid film, we added helium in small portions and recorded temperatures of each mentioned above transfer line section. Results of this measurement are presented in Table~\ref{tab:filling_line_overheating_He}. One can see that adding superfluid film somewhat cools the dissociator chamber and strongly heats up the Ac1, while the Ac2 and CI remain almost unchanged, with some weak cooling. Heating due to the re-condensing film increases with the film thickness, which leads mostly to the overheating of the Ac1 where obviously the major part of the helium vapor is condensed. The effect is saturated after adding 10~mbar$\cdot$dl meaning that the saturated helium film thickness is reached and adding extra helium just creates bulk liquid at the SC bottom. 

\begin{table}[]
    \centering
        \begin{tabular}{c|c|c|c|c}
            He            & Dissociator & Ac1 & Ac2 & CI  \\
            mbar$\cdot$dl & mK          & mK    & mK    & mK    \\
            0             & 644         & 265   & 249   & 146   \\
            5             & 603         & 357   & 241   & 152   \\ 
            10            & 612         & 437   & 246   & 143   \\
            20            & 613         & 436   & 240   & 138   \\
\\
        \end{tabular}
    \caption{The effect of the amount of He film on the temperatures of the H filling line sections with no RF discharge in the dissociator.}
    \label{tab:filling_line_overheating_He}
\end{table}

\begin{figure}
  \includegraphics[width=\columnwidth]{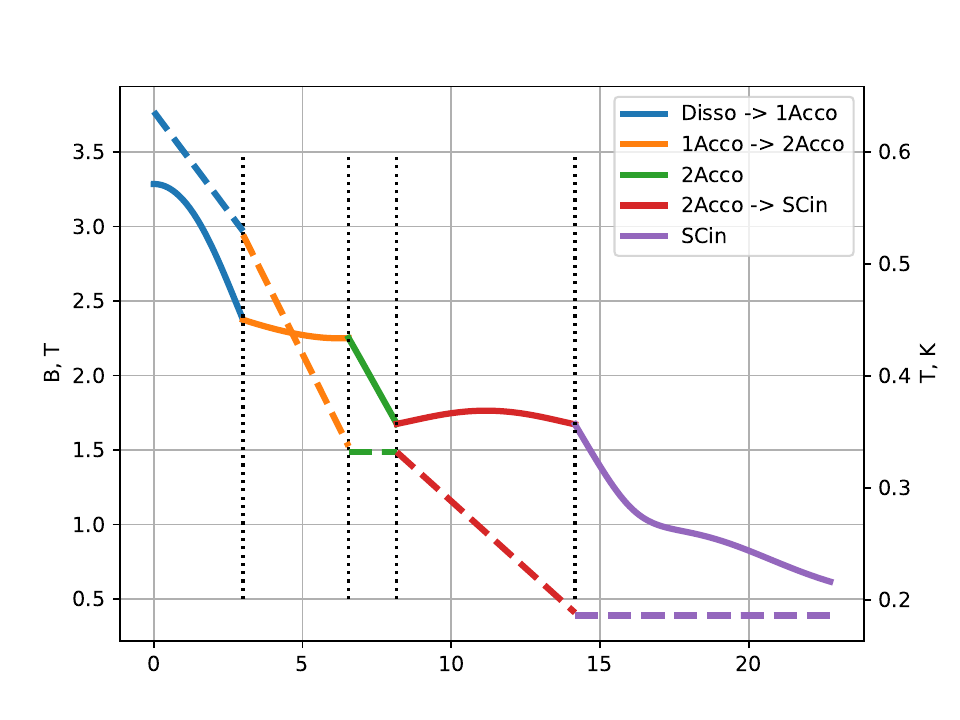}
  \caption{Magnetic field~(solid line) and temperature~(dashed line) along the path from dissociator to the end of the transfer line inside the Ioffe-Protchard trap.}
  \label{fig:BandTgradients}
\end{figure}

The magnetic field configuration is rather complex between the dissociator coil and the octupole magnets since the field decreases not only in magnitude but also changes the direction. The actual field profiles along the transfer line were evaluated using the BiotSavart program~\cite{BiotSavart} and used in the design of each section. The length of each section was done as short as possible, however taking into account that H atoms should experience on the average 1-2 sticking collisions with the wall. To increase the number of collisions, the 90$^{\circ}$ turns were included. This also prevents a direct access of ``hot'' atoms into the trapping cell. The turns as well help to fit all sections in a fairly small space available inside the dissociator coil and octupole magnets. In Fig.~\ref{fig:BandTgradients} we present a schematic plot of the magnetic field and temperatures changes across different section of the transfer line. One can see that the magnetic field drops by $\sim 1.5$~T between the dissociator and Ac2 and the temperature decreases from 600 to 320~mK on a short length of $\sim15$ cm. This provides effective separation of the \textit{lfs} and \textit{hfs} before the \textit{lfs} enter the CI, the coldest and longest part of the transfer line.

The last part of the transfer line~(CI) is made of a 90$^{\circ}$ turn at the top~(see Fig.~\ref{fig:Hsource_setup}) followed by a 3~mm~i.d. copper tube of the length $\approx 6$~cm. The lower end of the tube is covered by a short end cap spraying incoming atoms with the nearly 45$^{\circ}$ with respect to the trap axis. The end of the CI extends for about 5~cm below the upper flange of the trapping cell, in the region with magnetic field of $\approx 0.5$~T. This way we can load H gas into the trap at the level of roughly half of the trapping barrier and reduce the kinetic energy gained by the atoms when they settle in the trap. The incoming gas temperature of $\sim$130~mK is comparable with the magnetic energy difference of $\approx$0.2~K between the end of the CI and the trap bottom and a factor of $\sim 3$ smaller than the 540~mK trap barrier. This provides effective trapping of a large fraction of the incoming gas. The average number of sticking collisions in the CI is $\sim 3$, which is somewhat larger than needed. There is a danger of enhanced recombination on the walls of the CI, since the surface residence time increases exponentially with decreasing of temperature~\cite{Berkhout1986}. However, in this part of the line the gas is well polarized. The magnetic compression factor separating the \textit{lfs} from the \textit{hfs} is exp$\left(2\mu_B \Delta B/_B T\right)\approx 10^4$. We installed a heater on the CI outside the trapping cell, which can be used to warm up this section in order to decrease recombination. 

Analyzing thermal accommodation in different sections of the transfer line, we estimate the number of sticking collisions with the wall. On the surface the atoms may recombine or thermalize to the wall temperature and continue motion towards the trapping cell after the desorption. Recombination is not important on the first sections due to fairly high temperature. The sticking probability \textit{s}  of hydrogen atoms on the liquid helium surface vanishes at low energies and follows the law $s\approx 0.33\cdot T_g$~\cite{Berkhout1986}, where $T_g$ is the gas temperature. In the operating temperature range of our transfer line between Ac1 and CI the sticking probability varies between 0.12 and 0.04. Non-sticking collisions are mostly elastic and specular~(in case of sufficiently smooth surface)~\cite{Berkhout1993}. The number of sticking collisions in the circular tube of radius $r$ and length $l$ is $\sim s\cdot3/8 \left(l/r\right)^2$ \cite{Walraven1982}. Then, we estimate a required tube length for having one sticking collision as $l\sim r\sqrt{8/3s}$. For the transfer line sections Ac1, Ac2, and CI this gives the $l=2,~3$ and 3.5~cm accordingly. The lengths of first two stages are done close to this estimate. 

This consideration does not take into account the effect of the re-condensing of helium in the first two sections. Re-fluxing vapor works in a way similar to the old diffusion pump: atoms of the $^4$He vapor collide with H and push them towards the colder sections of the line. This effect was utilized for H gas compression in the research with \textit{hfs} and is known as HEVAC~(HElium VApor Compression)~\cite{BlueBible}. The quantitative evaluation of the HEVAC effect is not possible. But in general, it should make the transfer of atoms faster and decrease the probability of sticking. In terms of the thermalization efficiency this is compensated by collisions with the helium vapor.

\subsection{Sample cell for the trapped gas}
\begin{figure}
  \includegraphics[width=\columnwidth]{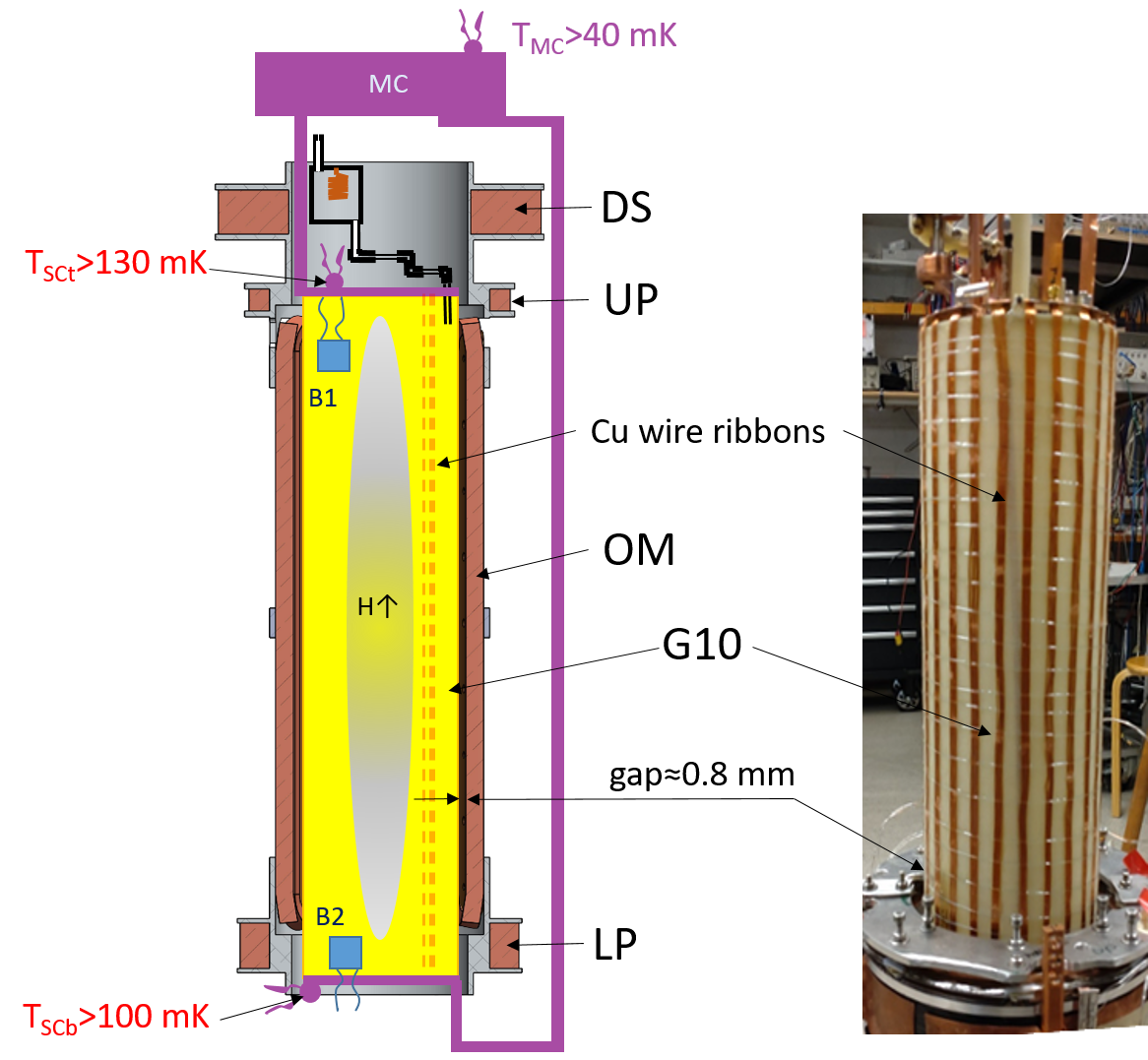}
  \caption{Left: schematic drawing of the sample cell inside the IPT. Right: Photo of the SC with the IPT lowered down.}
  \label{fig:TC}
\end{figure}

The hydrogen sample cell~(SC) is a cylinder made of G10~Glass Epoxy Laminate tube~\cite{Acculam} with inner diameter of 105~mm and wall thickness of 1.6~mm~(Fig.~\ref{fig:TC}). It fits inside the IPT with a $\approx0.8$~mm gap to the inner surface of the octupole system. Centering and control of the gap between SC and IPT is performed with 8~polyether ether ketone~(PEEK) rods screwed through the cores of the race track coils of the octupole magnets and pressing onto the SC tube with sharpened ends. Cooling of the SC is provided by the thermal links made of ribbons of 0.15~mm diameter insulated copper wire glued to the outer surface of the G10 tube. This helps reducing the heating of the SC by eddy currents during ramps of magnetic field. SC top and bottom end plates are manufactured from a 6~mm thick G10 sheet. Copper wire ribbons are glued to the copper plates at the SC top and bottom. Separate copper bars connect these plates to the mixing chamber of the DR. The bar for the SC bottom is placed outside the IPT. Temperature sensors and heaters are fixed to the top and bottom SC plates. This allows independent control of the SC top and bottom temperatures.

For the temperature measurements we used RuO$_2$ chips~\cite{Dale}. The sensors were installed on the main components of the dilution refrigerator: 1K~pot, Still, 0.1~K plate, Mixing chamber~(MC), as well as on the dissociator, Ac1, Ac2, CI, SC top and bottom flanges. For a control of the IPT temperature during current ramps, the sensors were attached to the OM, DS and LP coil mandrels.

In the future, we plan to implement two-photon spectroscopy of the 1S-2S transition for the diagnostics of the trapped gas. The laser system at 243~nm and luminescence detectors were not ready for this work, and we used calorimetric techniques for the characterization of the trap loading and decay of the H gas. The simplest way to do this is to monitor the recombination power of the atoms recombining on the inner SC walls. Accurate measurement of the power released in the SC is performed by active stabilization of the temperatures of the MC and the SC. Using two separate thermal links for the SC top and bottom plates required stabilizing them separately using dedicated sensors and thermometers. Changes of the voltages provided by the temperature controllers~(TC) of the SC top and bottom are detected during the SC loading and after switching off the H source. The absolute power released in the SC can be measured with $\sim0.1~\mu$W resolution which corresponds to the recombination rate of $N_{rec}\approx 3\cdot10^{11}$~s$^{-1}$. 

For reaching better sensitivity, we installed a bolometer at the bottom of the SC~(Fig.~\ref{fig:TC}). The bolometer has two temperature sensors located on different sides of a dielectric substrate. The sensors are carbon resistors made by painting Aquadag~(colloidal solution of carbon particles in water~\cite{SalonenBolometers}) between the thin electrodes sputtered on the substrate. We used a 0.1~mm thick sapphire sheet for the bolometer substrate with an area of $\approx2$~cm$^2$. The substrate is thermally coupled to the SC wall by a 50~$\mu$m thick superconductive wire of $\sim1$~cm length. Temperature sensors on both sides of the substrate provide the possibility of an absolute power measurement. For this, one Aquadag resistor is used as a heater using a high excitation current. A second resistor serves as a thermometer, measured with a very low excitation. This also allows stabilization of the substrate temperature at a given value avoiding uncertainties with the overheating effect of the Aquadag layer with respect to the helium film on it.

Bolometers are capable of detecting power changes at a pW level. However, they detect only a fraction of the total recombination heat equal to the ratio of the geometrical area of the bolometer to the area of the inner surface of the SC. This is $\approx1/700$ for our SC. Taking this into account, the bolometers are still a factor of 10-100 more sensitive than the temperature controllers described above. The presence of H gas provides an extra cooling channel via thermal accommodation of the gas~\cite{Goldman1986}. This effect depends on the gas density. As we shall see below, this makes interpretation of the bolometer signals rather complicated.

\section{Experimental results}
\subsection{Loading the sample cell with H gas}
In the first loading experiments, we charged the IPT coils to the currents which provide nearly the same trapping barrier for the axial and radial confinement. The currents which we fixed as nominal to operate the IPT safely with minimal risk of a quench were 120, 95, and 32~A for the octupole system, lower pinch, and upper pinch accordingly. We denote this configuration as the full ``100$\%$'' IPT charge. Later we will describe experiments where the IPT was charged to smaller fields. In all cases, the dissociator coil was charged to a nominal value of 95~A corresponding to $\approx3.5$~T. The full charge in the IPT provides a minimum trapping barrier of $\geq 0.8$~T~(0.54~K) in the so-called ``leakage points''~\cite{IPT_RSI} where the field of the octupole magnets is partially canceled by the field of the pinch coils. In the other regions the field is higher, reaching a maximum of $\approx 1.2$~T. Measurements were performed at SC temperatures in the range of 119...243~mK. The mixing chamber was stabilized at the lowest possible temperature which the DR could provide under a certain heat load from the SC, the transfer line components and the dissociator running the RF discharge. Typically the MC operated in the range 40-70~mK. The bolometer at the SC bottom was stabilized to 315~mK. 

\begin{figure}
  \includegraphics[width=\columnwidth]{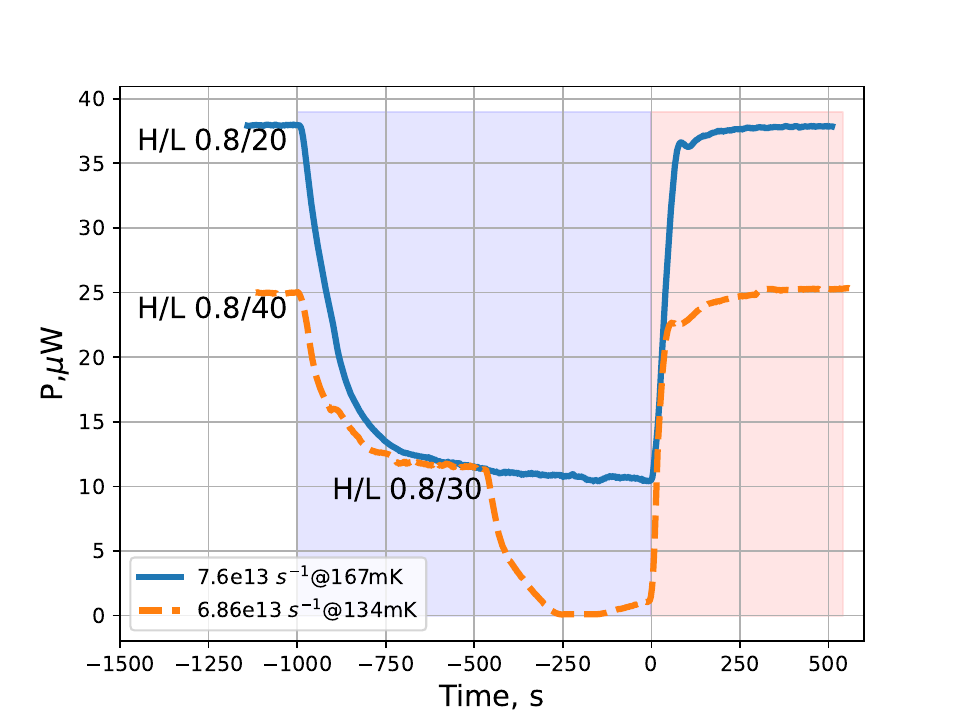}
  \caption{Results of the IPT accumulation and decay of the trapped gas for the SC temperatures of 167~(blue solid line) and 134~mK~(dashed brown line). The feedback power of the TC temperature controllers is plotted as a function of time. The discharge was turned on at $t\approx -1000$~s and turned off at $t=0$. RF pulse parameters are marked on the plot with H/L meaning RF on/off time in ms. Note that for measurement at 134~mK, at $\approx$ -500~s the dissociator L~(power OFF) time was decreased to 30~ms.}
  \label{fig:Accumulation}
\end{figure}

After stabilization of the SC temperature, we turned on the RF discharge in the dissociator and recorded voltages of the SC temperature controllers and the bolometer stabilization power. The power delivered to the dissociator was varied by changing the power of the RF oscillator by 6~dB or changing the duty cycle of its pulses. The average power absorbed in the dissociator running RF discharge was $\approx 0.5$~mW. Normally we used pulses of H/L~=~0.8/(20...40)~ms with ``H''$\equiv$ RF power ON and ``L''$\equiv$ RF power OFF times. After 10-15~minutes of loading we stopped the dissociator and monitored the decay of the trapped gas. In Fig.~\ref{fig:Accumulation} we presented results of such an experiment for the fully energized trap at the SC temperature of 134 and 167~mK. One can see that running the dissociator provides extra recombination power, which leads to a decrease of the temperature controller power required for stabilization of the SC. The effect is saturated after $\sim1000$~s of accumulation with a power change of $\approx$ 25~$\mu$W. Taking into account the recombination energy of $3.7\cdot 10^{-19}$~J~per~atom, we evaluate that a flux of $\approx 7\cdot 10^{13}~$~atoms/s is coming into the SC. For the $T_{SC}=134$~mK we first run the dissociator pulsing with H/L=0.8/40~ms, then increased the duty cycle by reducing L to 30~ms. This led to an increase of the detected atomic flux by about a factor of 2. Finally, the temperature controllers were not capable to keep stabilizing the SC since the recombination power exceeded the limit set by the temperature difference between the MC and SC ($\approx25 \mu$W in this case) with the dissociator OFF. The flux of atoms evaluated at 134~mK is about the same as for 167~mK.

In order to check that the observed change of the SC heating is related with the flux of atoms followed by the recombination, we run a control experiment when there was no molecular hydrogen uploaded into the dissociator. Running the discharge under the same conditions, we observed some heating of the dissociator and accommodators. No change of the heating power in the SC was detected within the noise, at the level of $\leq 1~\mu$W. In this regime some part of the superfluid helium film inside the dissociator is evaporated and re-condensed in the accommodators before reaching the SC. The large difference in the heating due to the hydrogen flux demonstrates a high efficiency of the transfer line and a proper design of the accommodators.

We present the temperatures of the dissociator and the other components of the transfer line for the dissociator ON/OFF conditions in Table~\ref{tab:filling_line_overheating_under_dis}. One can see that running the discharge, heats mostly the Ac1, somewhat less Ac2 and rather little the SC inlet.

\begin{table}[h!]
    \centering
        \begin{tabular}{c|c|c|c|c}
            Discharge   & Dissociator & Acco1 & Acco2 & SCin  \\
            OFF         & 592         & 364   & 221   & 154 \\ 
            ON          & 634         & 522   & 324   & 185 \\
            overheating & 41          & 158   & 102   & 31
        \end{tabular}       
    \caption{Temperatures~(mK units) of the H transfer line stages under discharge and its absence. }
    \label{tab:filling_line_overheating_under_dis}
\end{table} 

In order to find optimal operating parameters for getting the largest atomic flux, we varied the dissociator RF power, SC temperature and the magnetic field configuration. We found that the maximum flux is obtained at the lowest achievable SC temperature and largest tolerable dissociator power. These two parameters are related to each other. Increasing the dissociator power leads to  stronger heating, up to the limit of the refrigerator at the given temperature, like it is demonstrated for 134~mK in Fig.~\ref{fig:Accumulation}. At higher SC temperature, the refrigerator can absorb larger recombination power and we may increase dissociator power. However, we observed that increasing the RF power at SC temperatures above 150~mK leads to a decrease of the atomic flux. The data  presented in Fig.~\ref{fig:Accumulation} correspond to the optimal parameter configuration for reaching largest flux. We note that the temperature range above 150~mK is also not good for the trapping of H gas since the wall recombination of atoms escaping from the trap starts to be too slow for efficient evaporating cooling. 

The efficiency of the dissociator and the H transfer line can be evaluated by taking the ratio of the recombination heat released in the SC to the average RF power absorbed during discharge. At optimal RF power used for getting maximum flux, we measured that $\approx0.5$~mW is released in the dissociator. The recombination power in the SC measured at this condition is $\approx 25 \mu$W which corresponds to a dissociation efficiency of $\approx5\%$. This can be compared with the efficiency of 2$\%$ obtained by the Amsterdam group~\cite{VanRoijen1987} and 5$\%$ in MIT experiments working with a similar cryogenic dissociator~\cite{Killian1999}.

\subsection{Bolometer data}
\begin{figure}
  \includegraphics[width=\columnwidth]{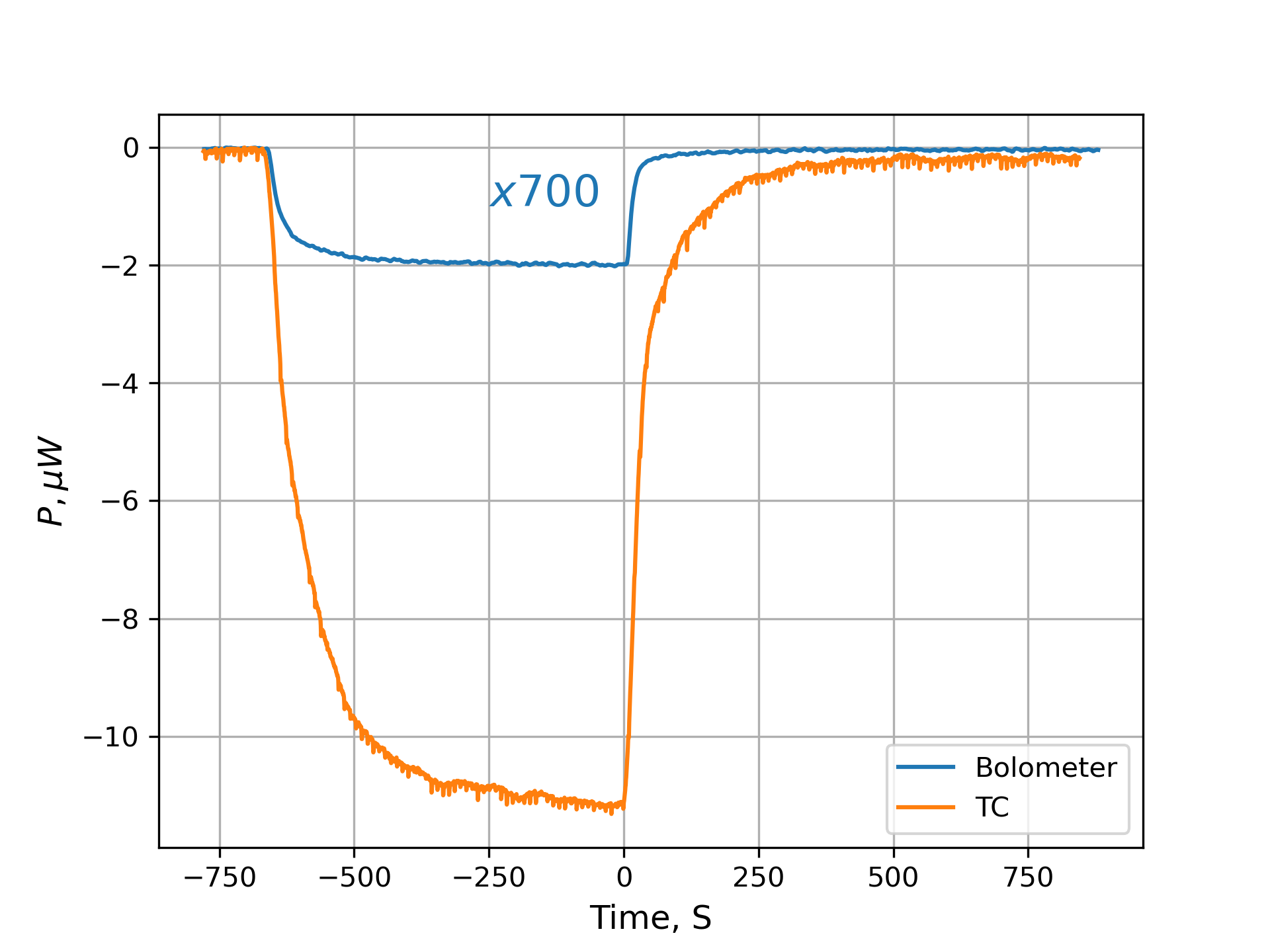}
  \caption{Comparison of the power changes detected by the SC temperature controllers and the bolometer at the SC bottom recorded during SC loading and decay at 134~mK. The bolometer signal is multiplied by the factor of 700 equal to the ratio of the SC inner wall area to the area of the bolometer surface.}
  \label{fig:TC_and_Bol}
\end{figure}

At SC temperatures below $\sim150$~mK the signals detected by the bolometer are very similar to that of the temperature controller. We present both signals at $T_{SC}=134$~mK in Fig.~\ref{fig:TC_and_Bol} for comparison. The heat detected by both methods originate from the recombination of the incoming atoms. It is well known that the recombination occurs at the surface of the experimental chamber and results in H$_2$ molecule in a highly excited ro-vibrational state. Only a small fraction of the recombination energy, $\leq1~\%$ is absorbed by the surface at the recombination site~\cite{Vasilyev1993}. The remaining  energy is distributed by the excited molecules to the rest of the walls. In this case, the power detected by the bolometer should be smaller than the total power detected by the temperature controller by the ratio of the geometrical area of the SC walls to the bolometer surface area. However, one can see in Fig.~\ref{fig:TC_and_Bol} that the ratio of powers $\sim 4000$ is much larger than the geometrical factor of $\approx700$ for our SC. This ratio increases to $\approx8000$ at 124~mK and decreases with increasing temperature. At T$_{SC} > 160$~mK it becomes equal to the geometrical factor. This observation implies that the bolometer detects a substantially smaller fraction of recombination power at low temperatures.

These results were obtained with the temperature of the bolometer (T$_b\approx315$~mK) being substantially higher than that of the SC walls. The above assumption of even distribution of the recombination energy between surfaces having different temperatures may not be correct. If the excited molecules have larger probability for relaxation in collisions with colder surface, then the colder walls will receive a larger fraction of the energy than the warmer parts. To our knowledge, the rate of the ro-vibrational relaxation in collisions with a wall was newer studied at so low temperatures. Our data indicate that below 150~mK the bolometers cannot be used for a reliable measurement of the absolute power released in recombination at the surface. However, this technique provides a more sensitive measurement of the power changes and can be used for the analysis above 160~mK.

\begin{figure}
  \includegraphics[width=\columnwidth]{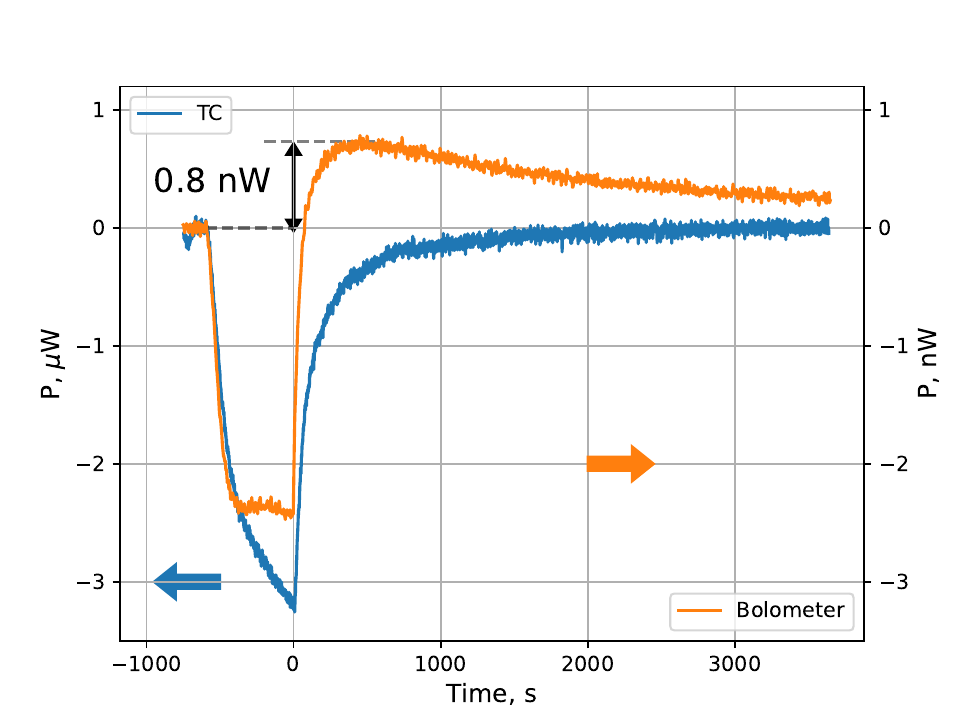}
  \caption{Comparison of the power changes detected by the temperature controllers and the bolometer for the SC temperature of 243~mK. Note the different vertical axes for the TC power~(left) and the bolometer~(right). A kind of ``overshooting'' behavior is seen in the bolometer signal after switching off the dissociator (at $t=0$). An extra cooling power of the order of 0.8-1~nW appears due to the thermal accommodation of H gas.}
  \label{fig:TC_and_Bol_240 mK}
\end{figure}

Results of the SC loading at a higher temperature T$_{TC}=243$~mK are presented in Fig.~\ref{fig:TC_and_Bol_240 mK}. One can see that after switching off the dissociator, the bolometer signal increases above its baseline before accumulation. There appears an extra cooling channel which leads to an increase of the power required for its stabilization. Such cooling is provided by the thermal accommodation of hydrogen gas present near the SC wall at the location of the bolometer. The gas has a temperature equal to that of the walls of the SC which is $\approx 72$~mK lower than the bolometer temperature. The cooling power $P_{acc}$ is proportional to the gas density near the bolometer $n_H$ and the temperature difference between the gas and the bolometer $\Delta T= T_{g} - T_{b}$:
\begin{equation}\label{eq:thermal_acc}
    P_{acc} = \frac{1}{2} n_H v_{th} A \alpha k_B\Delta T, 
\end{equation}
where \textit{A} is the bolometer surface area, $v_{th}$ is the thermal velocity of atoms, $k_B$ is the Boltzmann constant and $\alpha \approx 0.5\cdot T_g$ is the thermal accommodation coefficient~\cite{Berkhout1993}. Using this equation and the ``overshooting'' power of the bolometer $P_{acc}\sim1$~nW we evaluate that the density of the H gas near the bolometer is $\sim 8\cdot 10^{11}$~cm$^{-3}$. The magnetic field in the location of the bolometer is $\approx 0.9$~T and the magnetic compression factor for the gas located in the trap center is $\exp (\mu_B \Delta B/k_B T_{TC}) \approx 12$ for a temperature of 243~mK. As we shall see below, the maximum density of \textit{lfs} atoms in the trap center at 243~mK is about $1.2\cdot10^{12}$~cm$^{-3}$. Then, near the wall the density of \textit{lfs} should be $\sim10^{11}~$~cm$^{-3}$, much smaller than estimated from the bolometer cooling. Therefore, the gas which provides cooling cannot be in the \textit{lfs} state, but is most likely in the \textit{hfs}. At so high temperature of the SC this conclusion does not look surprising. The magnetic separation cannot effectively separate electron spin states and the recombination at the wall is slow enough to allow substantial density of H gas. The latter is also seen from very slow decay of the bolometer signal after switching off the dissociator.

For lower SC temperatures, bolometer data do not show a clear cooling effect. As expected, the cooling power due to the gas accommodation rapidly decreases with temperature due to the decrease of the gas density (\textit{lfs}) near the bolometer, which is an indication of a decoupling of the trapped gas from the wall. This is essential for the gas trapping and further evaporative cooling. 

Although, the bolometer is more sensitive for detection of the power changes in the SC during the loading and decay of H gas, the interpretation of its signal depends in a complicated way on the gas density and temperature. Therefore, we used the bolometer signal only for qualitative estimates, mostly at highest SC temperatures. For the absolute determination of the atomic flux and number of loaded atoms we used signals of the SC temperature controllers.

\subsection{Decay of the trapped gas}
\begin{figure}
  \includegraphics[width=\columnwidth]{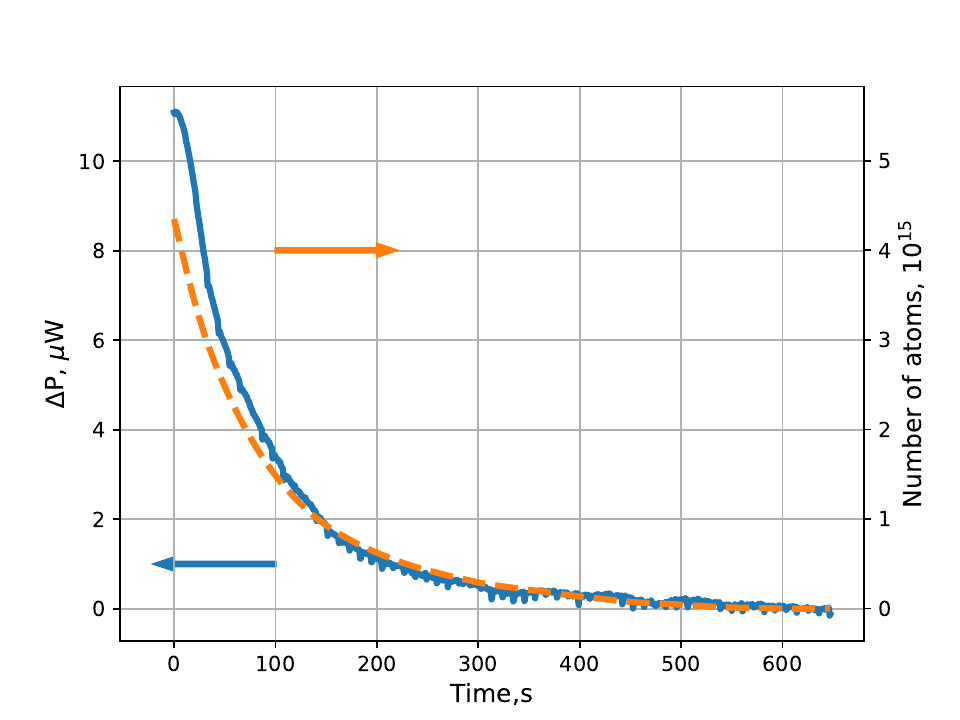}
  \caption{The decay of the trapped gas for the SC temperatures of 134~mK. Recombination power detected by the sample cell TC is shown (blue solid line). Number of atoms in the trap calculated from the integral of the TC signal is plotted with a dashed orange line.}
  \label{fig:Decay}
\end{figure}

Measuring the recombination rate with the SC temperature controllers after switching off the dissociator, when there is no incident flux, provides data about the decay of the accumulated sample and allows determination of the absolute number of atoms which are stored in the trap. This is done by integrating the TC power starting from the end of the decay. In Fig.~\ref{fig:Decay} we present results of such a measurement at $T_{SC}=140$~mK and after loading with a flux of $3\cdot10^{13}$~atoms/s. One can see that in the beginning of the sample decay we have $\approx 2\cdot 10^{15}$ atoms accumulated in the trap. 

After loading the trap and turning off the dissociator, the atoms having energies larger than the trap height $\varepsilon_t$ escape from the trap. The remaining atoms restore an equilibrium distribution with reduced temperature. This evaporative cooling mechanism is the main method used to cool H gas in magnetic traps~\cite{Masuhara88, Luiten96}. The atoms escaping from the trap collide with the sample cell walls. They can stick to the wall and recombine or desorb and return in the trap. Also, they can scatter elastically from the wall. For the evaporative cooling operating efficiently, it is required that the atoms recombine at the wall rather than return back into the trapping volume. This however, depends strongly on the SC wall temperature. Temperatures near 100~mK or lower are required to effectively eliminate evaporating atoms. In these experiments we are close to this condition at the lowest SC temperatures used. 

Relaxation to the \textit{hfs} states due to exchange and dipolar interactions is the second reason for the loss of the atoms. Various channels for the exchange and dipolar relaxation are analyzed and calculated in~\cite{Stoof1988}. During loading both \textit{lfs} hyperfine states  $c\equiv\mid F=1; M_{F}=0\rangle$ and $d\equiv\mid F=1; M_{F}=1\rangle$ are pushed into the trap. Soon after switching off the H source, \textit{c} atoms relax to the \textit{hfs} states due to rapid spin exchange in two-body collisions, escape from the trap and recombination on the SC walls. At this stage the number of atoms decreases most rapidly. Having no methods for direct measurement of the gas density and temperature, we cannot analyze what happens in our trap at this stage.

The rate of evaporative cooling slows down when the gas temperature decreases substantially below the trap depth. It was found in the experiments of the MIT group~\cite{DoyleThesis}, that the gas temperature is reduced to $\approx1/10$ of the trap height soon after stopping accumulation and remains stable at this level. At this stage the decay of the trapped \textit{d}-state gas is governed by a two-body dipolar relaxation. The rate constants for the dipolar relaxation was calculated in~\cite{Stoof1988} and experimentally determined by the MIT ~\cite{DoyleThesis} and Amsterdam ~\cite{vanRoijen1988} groups. It weakly depends on temperature and magnetic field and under the conditions of our experiments is in the range $1-2\times 10^{-15}$~cm$^3$/s.

\subsection{Effective volume and density of the trapped gas}
For characterization of the processes in the trapped gas we need to define the effective volumes of our trap. The evaluation of the effective volume is done by straightforward integration of the magnetic compression factor over the volume of the SC:
\begin{equation}\label{eq:Veff_classical}
    V_{1e} = \int e^{-\frac{\mu_B B(\textbf{r})}{k_B T}} d\textbf{r}.
\end{equation}

\begin{table}[]
    \centering
        \begin{tabular}{c|c|c|c}
            Trap depth ($\%$)    & 100        & 75         & 50 \\
            T$_g$, mK            &            &            &    \\
            60                   & 351/167    & 412/232    & 570/319  \\
            134                  & 631/342    & 813/453    & 1090/626  \\
            167                  & 763/416    & 970/546    & 1280/753  \\ 
            205                  & 904/499    & 1130/650   & 1444/889  \\
            243                  & 1033/589   & 1272/740   & 1589/1014  \\
\\
        \end{tabular}
    \caption{Effective volumes for H gas in our SC $V_{1e}/V_{2e}$ in cm$^{3}$ for different temperatures (left column) and magnetic field configurations: 100$\%$, 75$\%$, and 50$\%$of the full IPT charge.} 
    \label{tab:Veff}
\end{table}

Knowing the effective volume and total number of atoms $N$ one can find density of the gas in the trap center as $n_0~=~N/V_{1e}$. Since we do not have yet any method to measure the gas temperature, an accurate determination of the effective volume is not possible. Assuming that during the accumulation the gas temperature is close to that of the SC wall, we obtain a lower bound for the gas density in the trap bottom of $n_0\approx 3\cdot10^{12}$~cm$^{-3}$ in the beginning of the decay shown in Fig.~\ref{fig:Decay}. The gas can be colder during the loading process due to evaporative cooling and also due to the fact that we deliver it into the middle of the trapping potential where the end of the cell inlet line is extended. In the other extreme case when the gas temperature is 1/10 of the trap depth, $\approx60$~mK, the effective volume is nearly factor of 2 smaller and we would get $n_0\approx 6\cdot10^{12}$~cm$^{-3}$.

We may also make a rough estimate of the steady state density in the end of the gas accumulation using the known value of the dipolar relaxation rate constant $G_{dd}$ and assuming that the atoms are mostly accumulated in the doubly polarized \textit{d}-state, which is known from other experiments ~\cite{Masuhara88, Luiten96}.  In the steady state, the incoming flux $\Phi$ is balanced by the losses due to the relaxation and subsequent recombination in the trapped gas having the rate $G_{dd}n^2$. Integrating over the density distribution, we may write:
\begin{equation}\label{eq:Decay_equation}
    \frac{dN}{dt} \equiv \frac{dn_0}{dt}V_{1e} = \Phi - 2G_{dd} n_0^2 V_{2e}, 
\end{equation}
where the second term in the right hand side is the loss due to relaxation and
\begin{equation}\label{eq:Decay_equation2}
   V_{2e}=\int e^{-\frac{2\mu_B B(\textbf{r})}{k_B T}} d\textbf{r} 
\end{equation}
is the second order effective volume for the two body process of dipolar relaxation~\cite{Luiten96}. We present calculations of the effective volumes for the range of temperature and magnetic fields used in this experiments in Table~\ref{tab:Veff}.

In a steady state we set the time derivative to zero in Eq.~\ref{eq:Decay_equation} and get:
\begin{equation}\label{eq:Density_from_flux}
    n_0= \left(\frac{\Phi}{2G_{dd}V_{2e}}\right)^{1/2}.
\end{equation}
The data for the incoming flux $\Phi$ are obtained from the TC signal as explained above. Then taking the incoming flux of $3\cdot10^{13}$~atoms/s, $V_{2e}\approx400$ cm$^3$ evaluated for 134~mK, and $G_{dd}=1.5\cdot10^{-15}$~cm$^{3}$/s taken from Ref.~\cite{Stoof1988} we obtain $n_0=3.6\cdot10^{12}$~cm$^{-3}$. A more detailed consideration of relaxation processes which occur at the accumulation stage is presented in Appendix A. In a steady state we obtained the densities of the \textit{lfs} hyperfine state $n_c\approx0.7 \cdot10^{12}$ and $n_d\approx3.6 \cdot10^{12}$~cm$^{-3}$. Comparing the total density $n_d + n_c\approx4.3 \cdot10^{12}$~cm$^{-3}$ with the above estimate obtained from the integrated TC signal, we can conclude that the trapped gas temperature is somewhere between 60 and 134~mK and the gas is cooled during the loading process below the SC wall temperature. This is what we expected to reach with the technique of loading into the middle of the trapping potential.

After loading the trap, we attempted to perform forced evaporative cooling of the trapped gas by lowering the field of the IPT coils. This procedure is technically fairly simple and can be done with the rate sufficient for effective evaporative cooling. However, the eddy current heating of the SC due to the field ramp is comparable with the recombination heat of escaping atoms already in the beginning of the ramp. The number of trapped atoms decreases by orders of magnitude in the evaporative cooling cycle and cannot be detected calorimetrically.

\subsection{Loading at various trap depths and SC temperatures}
\begin{figure}
  \includegraphics[width=\columnwidth]{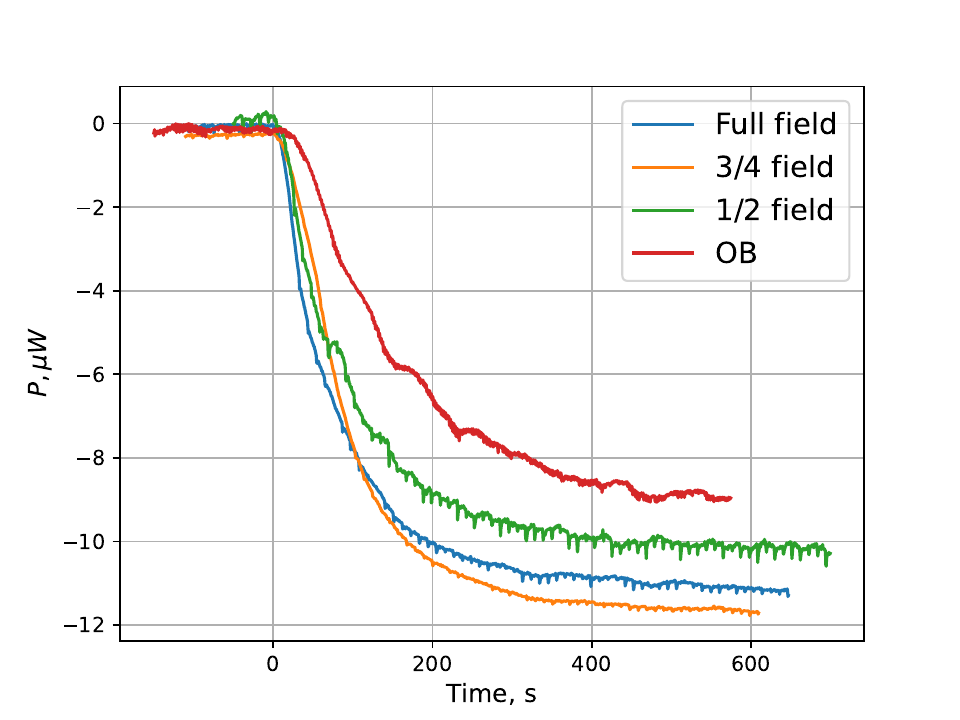}
  \caption{TC power changes during accumulation at the $T_{SC}=134$~mK and various configurations of the magnetic field in the IPT: 100$\%$, 75$\%$, 50$\%$, and for the configuration~``0$\%$''~(open bottom). }
  \label{fig:var_B}
\end{figure}

\begin{figure}
  \includegraphics[width=\columnwidth]{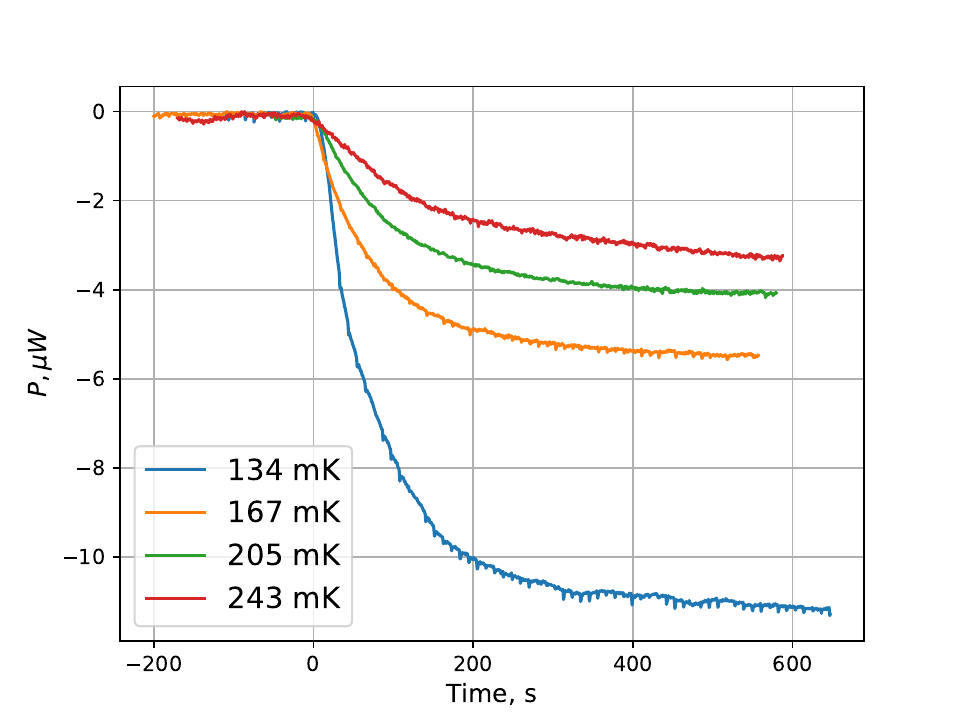}
  \caption{TC power changes during accumulation at the nominal magnetic field in the IPT~(100$\%$) and various sample cell temperatures.}
  \label{fig:var_T}
\end{figure}

We performed a series of measurements with the IPT coils energized to 100$\%$, 75$\%$, 50$\%$ and for the configuration ``0$\%$'' when the DS and UP pinch were energized at 100\%
while all other coils fully discharged (zero field in IPT), so-called open bottom configuration~(OB). In the latter case the trap is open at the bottom while there is $>0.6$~T field in its upper part. The atoms are not confined magnetically at the lower part of the SC and freely interact with the SC walls. In Fig.~\ref{fig:var_B} we present results of the temperature controller signal during accumulation of the H gas for the above mentioned magnetic field configurations in the IPT. At each configuration we performed a series of measurements for various SC temperatures in the range $T_{SC}=134-255$~mK. In Fig.~\ref{fig:var_T} we present results of the accumulation for some values of temperature  in this range and fully charged IPT.

One can see that decreasing magnetic field has a weak effect on the incoming flux into the SC. The TC signals in Fig.~\ref{fig:var_B} are saturated at smaller power changes. However, this decrease reaches at most 25$\%$ for the case of the trap open at the bottom. This indicates that the operation of the dissociator and H transfer line does not depend on the magnetic field in the IPT coils which provide magnetic confinement for \textit{lfs} of H gas. All H gas entering the SC eventually recombine when the accumulation is saturated. The total power released in recombination solely depends on the incoming flux which is nearly the same irrespective of the field in the IPT.

\begin{figure}
  \includegraphics[width=\columnwidth]{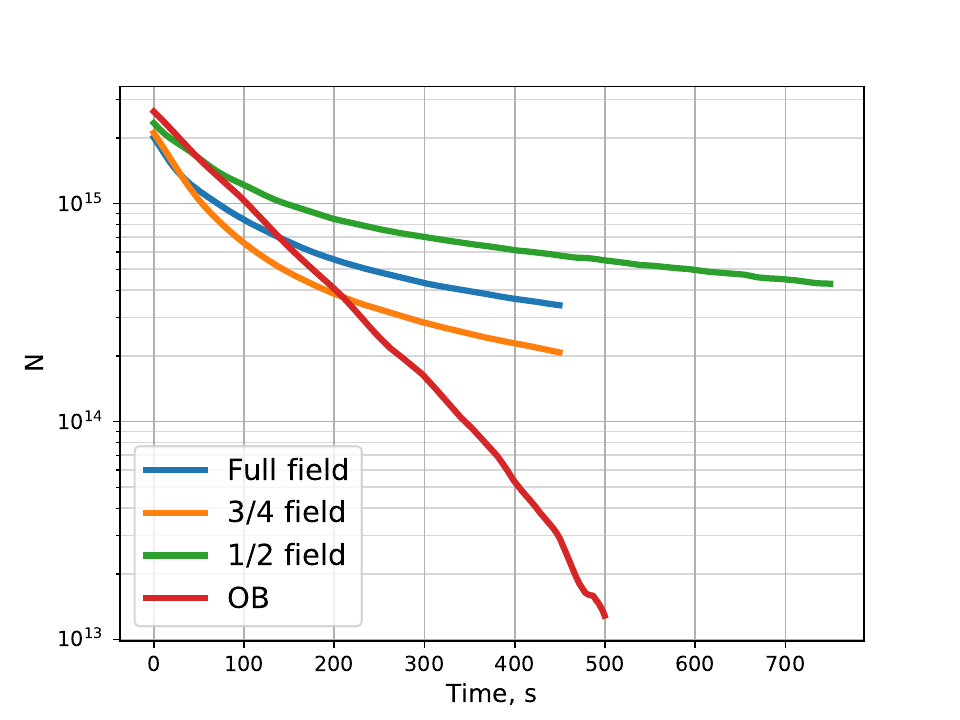}
  \caption{Total number of stored H atoms after accumulation in various magnetic field configurations and SC temperature of 134~mK.}
  \label{fig:N_for_var_B}
\end{figure}

However, at decreased trapping field the coupling of the atoms to the walls is increased, which should lead to a faster recombination. Since the total recombination rate is nearly the same, this means that the steady state density at lower trapping fields should be smaller. To verify this we integrated the TC signals at the decay stage. This provides the total number of stored atoms in the end of the accumulation. Results are presented in Fig.~\ref{fig:N_for_var_B}.

The trapped number of atoms somewhat decreases for reduced field, from $3.5\cdot10^{15}$ at fully charged trap down to $2\cdot10^{15}$ for the trap with the open bottom. Now, if we take the values of the effective volume $V_{1e}$ from the Table~\ref{tab:Veff}, we get the following starting densities: $3.5\cdot10^{12}$~cm$^{-3}$ for the full field;  $2.3\cdot10^{12}$~cm$^{-3}$ for $75\%$; $1.7\cdot10^{12}$~cm$^{-3}$ for $50\%$; $1.2\cdot10^{12}$~cm$^{-3}$ for $0\%$. 

Comparing the loading curves at various temperatures in Fig.~\ref{fig:var_T}, we can see that the flux of atoms decreases for higher SC temperature, a feature which we already discussed in Sec.~5.1. At the highest temperature of 243~mK, we have not reached steady state after 10 minutes of accumulation. Filling of the SC proceeds much more slowly due to a larger effective volume and reduced incoming flux.
 
\section{Summary and conclusions}
\label{sec:application}
We have demonstrated successful operation of our cold H source for loading a large magnetic trap, with a flux of atoms of  $7\cdot10^{13}$~atoms/s and with an overall $\sim5\%$ efficiency of the dissociation plus thermal accommodation to $\sim130$~mK. The heat released in the dissociator running the RF discharge did not exceed 0.5~mW which is fairly small value even for modest power dilution refrigerator. The RF power can be increased by an order of magnitude for getting a larger flux of atoms. In the present work this does not help because of too a weak thermal link between the SC and mixing chamber of the DR. Increasing the flux of atoms just leads to an overheating of the SC. The enhancement of the thermal link by an order of magnitude can be easily done. This will allow a substantial increase of the atomic flux. We estimate that with the technique described in this work atomic fluxes approaching $10^{15}$~atoms/s can be reached. The thermal link improvement will also allow operation at temperatures of the SC below 100~mK. This will ensure better decoupling of the trapped gas from the walls and faster removal of the atoms escaping from the trap which is important for efficient evaporative cooling of the gas.

Our H source provides nearly the same flux of atoms into the sample cell when the trapping field is reduced to zero. This can be used for a wide range of other applications where an intense cold beam of H is required, not using a strong and expensive magnetic trap. At 130~mK the average thermal velocity of H atoms is $\approx 50$~m/s. This velocity range is sufficient for the observation of the Gravitational Quantum States as planned by the GRASIAN collaboration \cite{Killian2024}. Such slow atoms can be transported over substantial distances using a tube lined with a superfluid helium film due to the enhanced reflectivity from the helium surface~\cite{Berkhout1993, Yu1993}. They can be converted in to a beam using a parabolic mirror \cite{Luppov1993} or focused to a point \cite{Berkhout1993}. These manipulations in the phase space can be done with minimal losses of the atomic number. Releasing the beam into a vacuum outside the cryostat, can be done after solving a problem with the superfluid helium film flowing towards the warm regions. There are ready solutions for this, based on a special film cutters~\cite{Kaufman1993, HeCsCutter1, HeCsCutter2}. Further slowing of the H beam using a Zeeman decelerator~\cite{Merkt2008} should be fairly easy. Starting with 50~m/s, two deceleration stages will be enough for a full stop of the beam.

\section*{Acknowledgments}
This project was supported by the Jenny and Antti Wihuri foundation.
FN and PY acknowledge support from IEA QRECH 2021-2022 and IRP GRASIAN 2024-2028. The work of PC was supported by the European Research Council (grant 818053-Mu-MASS) and the Swiss National Science Foundation (grants 197346 and 219485).

\section*{Author contributions}
All authors contributed to the concept and the design of the experimental setup. Setting up and running the experiment were performed by AS, JA, SD, and SV. Data collection and processing was done by AS, JA and SV. The first draft of the manuscript was written by AS and SV. All authors commented on previous versions of the manuscript and approved its final version. 

\subsection{Data Availability Statement}
 The datasets generated and/or analyzed during the current study are available from the corresponding author on request.

\begin{appendices}
\section{} 
\begin{figure}
    \includegraphics[width=\columnwidth]{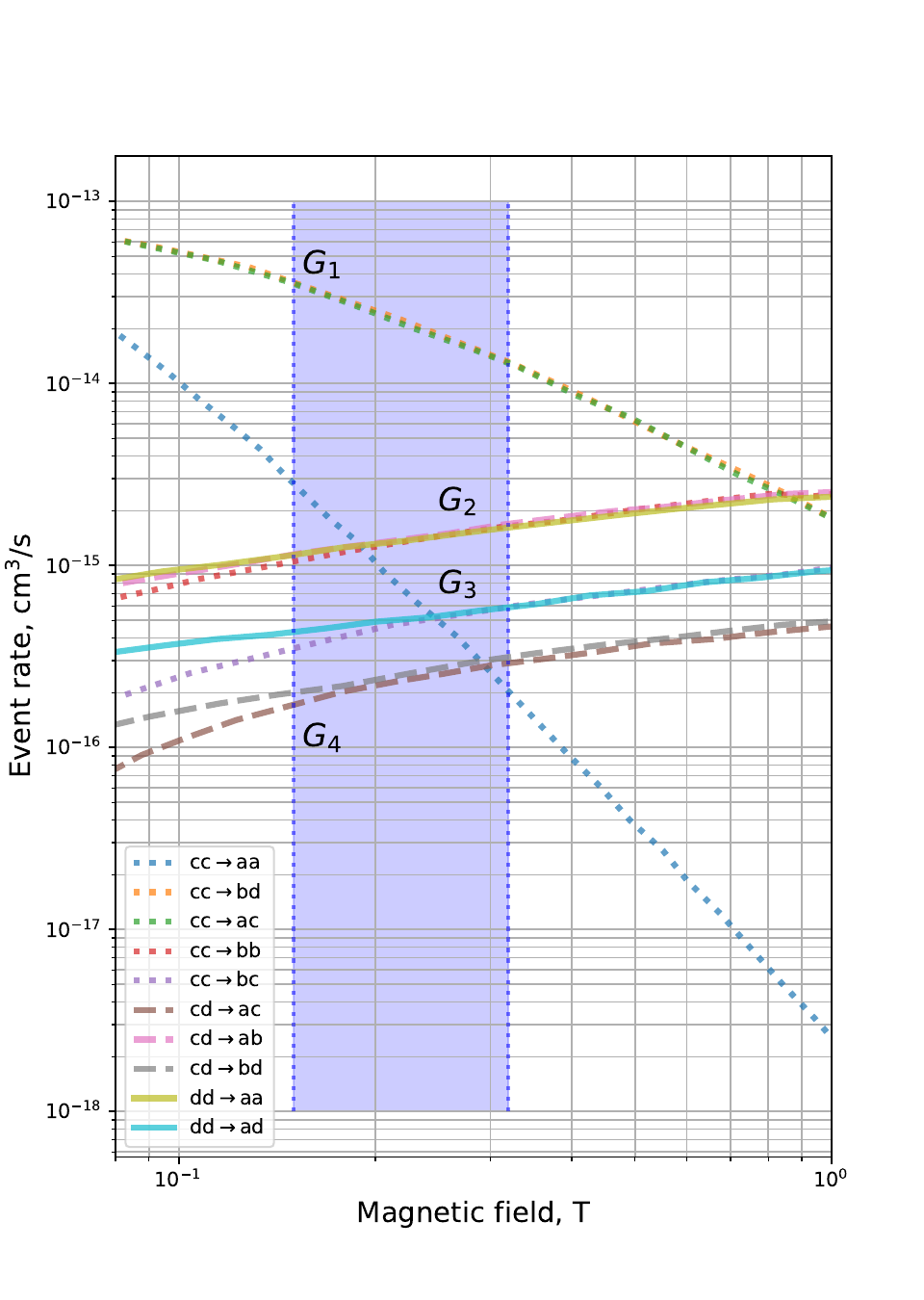}
    \caption{Plot of the relaxation rate constants for the processes relevant for our case. Adapted from ref.~\cite{Stoof1988}}
    \label{fig:RelRates}
\end{figure}

Relaxation processes which may occur in a magnetically trapped gas due to the exchange and dipolar interactions in binary collisions were analyzed in ref.~\cite{Stoof1988}. Collisions of pairs of atoms of all four hyperfine states were considered in this work. In the following consideration, we restrict ourselves to the binary collisions of the atoms in the two \textit{lfs} hyperfine states $c\equiv\mid F=1; M_{F}=0\rangle$ and $d\equiv\mid F=1; M_{F}=1\rangle$ which may however, after relaxation result to any pair of the four, including the \textit{hfs} states  $a\equiv\mid F=0; M_{F}=0\rangle$ and $b\equiv\mid F=1; M_{F}=-1\rangle$. We present relevant a relaxation rates in Fig.~\ref{fig:RelRates} as a function of the magnetic field at zero temperature. The field of interest between 0.15 and 0.32~T is outlined. This region corresponds to the magnetic field experienced by the trapped atoms for 100$\%$ trap charge. We follow the notations of Ref.~\cite{Stoof1988} for the relaxation channels between the pairs of atoms, e.g. \textit{cc-ac} means collision of two \textit{c}-state atoms resulting to atoms in the \textit{a} and \textit{c} states with a corresponding rate constant denoted as $G_{ccac}$. Our next assumption is that the atoms of the \textit{hfs} states \textit{a} and \textit{b} resulting from relaxation events leave the trap and recombine of the SC walls. We assume that the hydrogen source provides equal fluxes of atoms $\Phi_c=\Phi_d=\Phi/2$ of the \textit{c} and \textit{d} states. Then, the rate equations for the relaxation loss rate for bulk densities $n_c(\textbf{r})$ and $n_d(\textbf{r})$ of the \textit{c} and \textit{d} states will include the following terms:

\begin{equation}\label{eq:Relaxation_equations1}
\begin{split}
\frac{dn_c}{dt} = &- n_c^2\left(2G_{ccaa}+2G_{ccbd} \right. \\
                   &\left. +G_{ccac}+G_{ccbc}+2G_{ccbb} \right) \\
                   &-n_c n_d \left(G_{cdab}+G_{cdbd}\right) 
\end{split}
\end{equation}

\begin{equation}\label{eq:Relaxation_equations2}
    \begin{split}
   \frac{dn_d}{dt} = & - n_d^2\left(2G_{ddaa}+G_{ddad}\right)   \\ 
                     & - n_c n_d \left(G_{cdab}+G_{cdbd}\right) \\ 
                     & + n_c^2G_{ccbd}
   \end{split}
\end{equation}
As it is seen in the Fig.~\ref{fig:RelRates}, for the magnetic field region where our trapped gas is located (0.15-0.32~T), the relevant relaxation rate constants merge into four major values: $G_{ccbd}\approx G_{ccac}\approx 2\cdot 10^{-14}$~cm$^{-3}$/s, $G_{ddaa}\approx G_{cdab}\approx G_{ccbb}\approx1.5\cdot 10^{-15}$~cm$^{-3}$/s, $G_{ddad}\approx G_{ccaa}\approx G_{ccbc}\approx5\cdot 10^{-16}$~cm$^{-3}$/s, $G_{cdbd}\approx G_{cdac}\approx 2.5\cdot 10^{-16}$~cm$^{-3}$/s, which we denote as $G_1$, $G_2$, $G_3$, and $G_4$ accordingly. Then, integrating Eqs.~\ref{eq:Relaxation_equations1} and \ref{eq:Relaxation_equations2} over the volume of the trap we obtain a set of equations~(see also Eq.~\ref{eq:Decay_equation} and definition of the effective volumes in Sec.5.4):
\begin{equation}\label{eq:Relaxation_equations3}
    \begin{split}
        \frac{d n_{c0}}{dt} V_{1e} = \frac{\Phi}{2} - [& n_{c0}^2(3G_3+3G_1+2G_2) \\ 
                                                  & + n_{c0} n_{d0}(G_2+G_4)  ]  V_{2e}
    \end{split}
\end{equation}
\begin{equation}\label{eq:Relaxation_equations4}
    \begin{split}
        \frac{d n_{d0}}{dt}V_{1e}  = \frac{\Phi}{2} - [&n_{d0}^2(2G_2+G_3) \\ 
                                                  & + n_{c0} n_{d0} (G_2+G_4) \\
                                                  &  - n_{c0}^2G_1 ] V_{2e}
    \end{split}
\end{equation}
for the densities $n_{i0}$ in the trap center.
In the steady state during accumulation we set time derivatives to zero. Subtracting the second equation from the first, we get
\begin{equation}\label{eq:c/d_ratio}
\small{
  \frac{n_{c0}^{s}}{n_{d0}^{s}} = \left(\frac{2G_2+G_3}{4G_1+2G_2+3G_3}\right)^{1/2}},
\end{equation}
which gives $n_{c0}^s /n_{d0}^s\approx 0.2$ after substituting the numerical values of the rate constants.
Then, using this density ratio and summing the equations \ref{eq:Relaxation_equations3} and \ref{eq:Relaxation_equations4} after simple algebra, we evaluate the steady state density of the \textit{d}-state in the trap center:

\begin{equation}\label{eq:d_steady}
  n_{d0}^{s}\approx 1\cdot10^7 [cm^{-3}\cdot s]^{1/2} \left(\frac{\Phi}{V_{2e}}\right)^{1/2}
\end{equation}
For the effective volume $V_{2e}\approx 390$~cm$^{3}$ evaluated at 134~mK and a total incoming flux of $3\cdot10^{13}$~atoms/s we get the values of the steady state densities during accumulation $n_{c0}^{s} \approx 7\cdot10^{11}$~cm$^{-3}$ and $n_{d0}^{s} \approx 3.6\cdot10^{12}$~cm$^{-3}$.

\end{appendices}


\bibliography{H_source}


\end{document}